\documentclass[twoside,leqno,twocolumn]{article}  
\usepackage{ltexpprt}

\usepackage[noend, noline ,boxruled]{algorithm2e}
\usepackage{graphicx}
\usepackage{color}
\usepackage{amsmath}
\DeclareMathOperator*{\argmax}{\arg\!\max} 
\allowdisplaybreaks 

\usepackage[font=scriptsize,labelfont=bf]{caption}

\usepackage[compact]{titlesec}
\titlespacing{\section}{2pt}{3pt}{2pt}
\titlespacing{\subsection}{2pt}{3pt}{2pt}
\titlespacing{\subsubsection}{2pt}{3pt}{2pt}
\begin{document}

\title{\Large Mining Block I/O Traces for Cache Preloading with  Sparse Temporal Non-parametric Mixture of Multivariate Poisson\thanks{This 
work was done in collaboration with NetApp, Inc. }}
\author{Lavanya Sita Tekumalla\thanks{Indian Institute of Science} \\
\and 
Chiranjib Bhattacharyya\footnotemark[2]
}
\date{}

\maketitle

 \begin{abstract}
Existing caching strategies, in the storage domain, though well suited to exploit short range spatio-temporal patterns, are unable to leverage long-range motifs for improving hitrates. Motivated by this, we investigate novel Bayesian non-parametric modeling(BNP) techniques for count vectors, to capture long range correlations for cache preloading, by mining Block I/O traces. Such traces comprise of a sequence of memory accesses that can be aggregated into high-dimensional sparse correlated count vector sequences. 

While there are several state of the art BNP algorithms for clustering and their temporal extensions for prediction, there has been no work on exploring these for correlated count vectors. Our first contribution addresses this gap by proposing a DP based mixture model of Multivariate Poisson (DP-MMVP) and its temporal extension(HMM-DP-MMVP) that captures the full covariance structure of multivariate count data. However, modeling full covariance structure for count vectors is computationally expensive, particularly for high dimensional data. Hence, we exploit sparsity in our count vectors, and as our main contribution, introduce the ”Sparse DP mixture of multivariate Poisson(Sparse-DP-MMVP)”, generalizing our DP-MMVP mixture model, also leading to more efficient inference. We then discuss a temporal extension to our model for cache preloading.

We take the first step towards mining historical data, to capture long range patterns in storage traces for cache preloading. Experimentally, we show a dramatic improvement in hitrates on benchmark traces and lay the groundwork for further research in storage domain to reduce latencies using data mining techniques to capture long range motifs.

\end{abstract}

\def\wl{\tilde{l}}
\def\wk{\tilde{k}}
\def\wt{\tilde{t}}
\def\wa{\tilde{a}}
\def\wb{\tilde{b}}
\def\mt{{\mathbf T}}
\def\mz{{\mathbf Z}}	

 \section{Introduction}
 \label{intro}


Bayesian non-parametric modeling, while well explored for mixture modeling of categorical and real valued data, has not been explored
for multivariate count data. We explore BNP models for sparse correlated count vectors 
to mine block I/O traces from enterprise storage servers for \textit{Cache Preloading}.

Existing caching policies in systems domain, are either based on eviction strategies of removing the least relevant data 
from cache (Ex: Least Recently Used a.k.a LRU) or read ahead strategies for sequential access patterns.
These strategies are well suited for certain types of workloads where nearby memory accesses are correlated in extremely short 
intervals of time, typically in milli-secs. However, often in real workloads, we find correlated memory accesses 
spanning long intervals of time (See fig 1), exhibiting no discernible correlations over 
short intervals of time (see fig 2).

There has been no prior work on analyzing trace data to \textit{ learn long range access patterns} for {\bf predicting future accesses.}
We explore caching alternatives to automatically learn long range spatio-temporal 
correlation structure by analyzing the trace using novel BNP techniques for count data, and exploit it to pro-actively preload data into cache and 
improve \emph{hitrates}.
\begin{figure}[b!]
\centering \includegraphics[height=1.5in,width=3in]{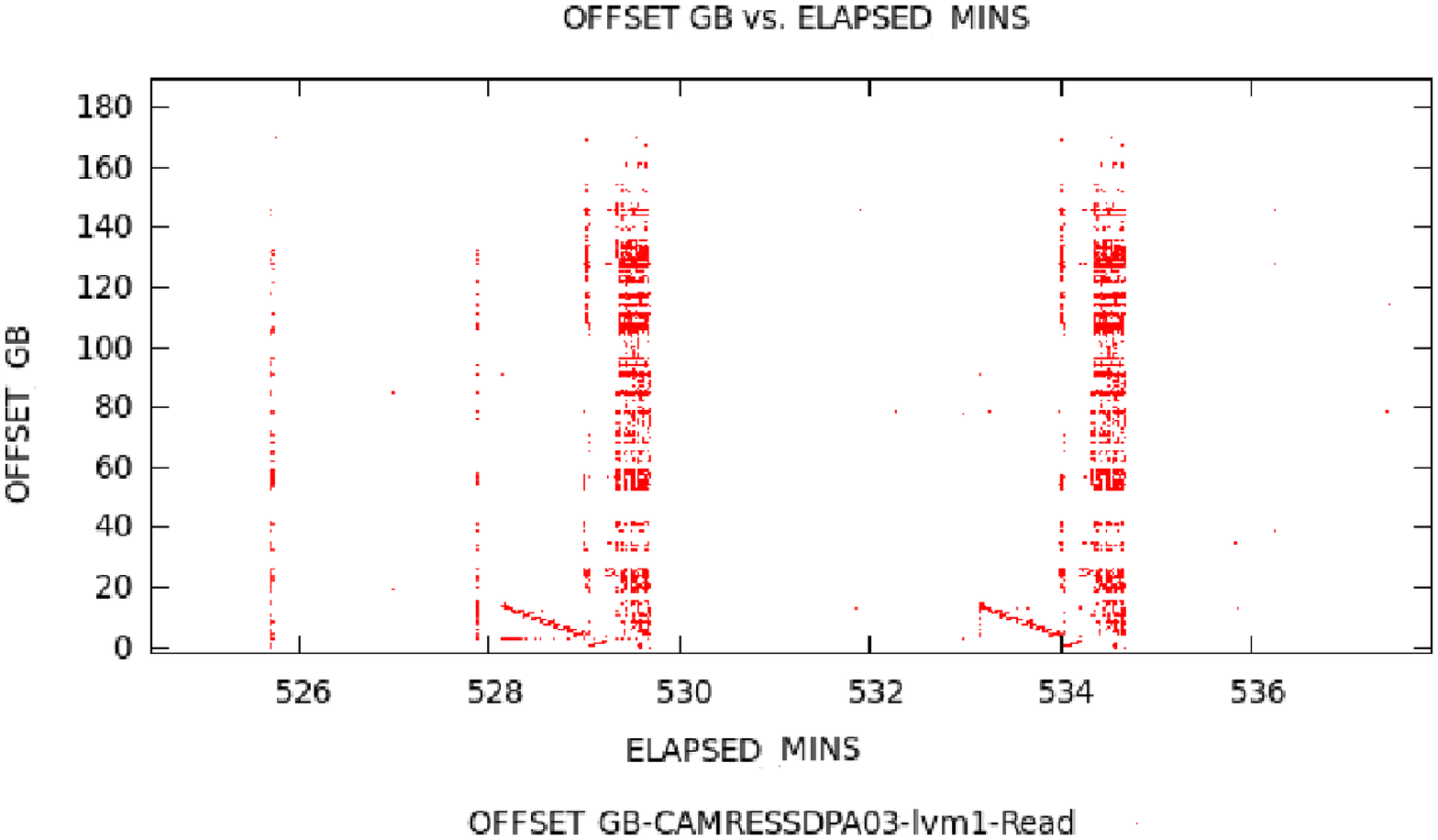}  
\centering \caption{LBA accesses vs time plot for MSR trace(CAMRESWEBA03-lvm2): A Zoomed out(coarser view) shows repeating  
access patterns over a time span of minutes.}  
\label{fig:intropica}  
\centering \includegraphics[height=1.5in,width=3in]{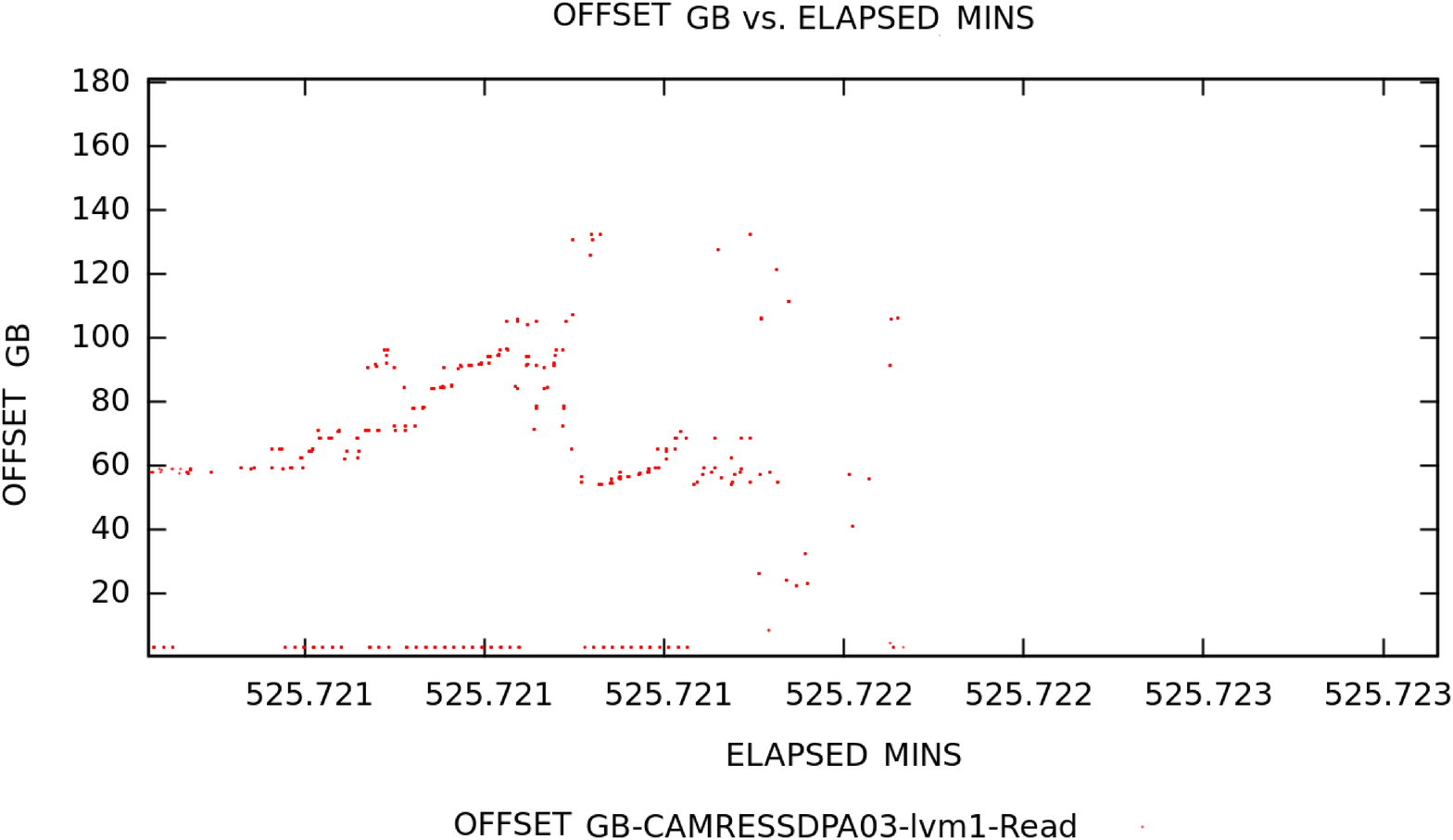}  
\centering \caption{A zoom-in access level view of this trace, into one of these patterns, over a time span of $\sim$ 100 milli-secs,
does not show any discernible correlation. Also note the sparsity of data; during any interval of time, only a small 
subset of memory locations are accessed 
\label{fig:intropicb}}
\end{figure}  

{\bf Capturing long range access patterns in Trace Data: }
Block I/O traces comprise of a sequence of memory block access requests (often spanning millions per day). 
We are interested in mining such traces to capture spatio-temporal 
correlations arising from repetitive long range access patterns (see fig \ref{fig:intropica}). 
For instance, every time a certain file is read, a similar sequence of accesses might be 
requested.

We approach this problem of capturing the long-range patterns by taking a 
more aggregated view of the data to understand longer range dependencies. We partition both the memory and time into 
discrete chunks and constructing histograms over memory bins for each time slice (spanning several seconds) to get a 
coarser view of the data. Hence, we aggregate the data into a sequence of count vectors, one for each time slice, 
where each component of the count vector records the count of memory access in a specific bin 
(a large region of memory blocks) in that time interval. Thus, a trace is transformed into a sequence of count vectors.

Thus, the components of count vector instances after aggregation are correlated within each instance since memory access 
requests are often characterized by \textit{spatial correlation}, where adjacent regions of memory are likely 
to be accessed together. This leads to a rich covariance structure. Further, due to the long range temporal dependencies in
access patterns (fig \ref{fig:intropica}), the  sequence of aggregated count vectors are also temporally correlated.
Count vectors thus obtained by aggregating over time and space are also sparse, where only a small portion of 
memory is accessed in any time interval (fig 2). Hence a small subset of count vector dimensions 
have significant non zero values. Modeling such sparse correlated count vector 
sequences to understand their spatio-temporal structure remains unaddressed. 

{\bf Modeling Sparse Correlated Count Vector Sequences: }
A common technique for modeling temporal correlations are Hidden Markov Models(HMMs) which are mixture model extensions
for temporal data. Owing to high variability in access patterns inherent in storage traces,  
finite mixture models do not suffice for our application since the number of mixture components varies often based on
the type of workload being modeled and the kind of access patterns. BNP techniques 
address this issue by automatically adjusting the number of mixture components based on complexity of data. 
Non-parametric clustering with Dirichlet Process(DP) mixtures  and their temporal variants have been 
extensively studied over the past decade \cite{DP}, \cite{HDP}. However, to the best of our knowledge
we are not aware of such models in the context of count data, particularly temporally correlated sparse count vectors.


Poisson distribution is a natural prior for counts and the Multivariate Poisson(MVP) for correlated count vectors. 
However, owing to the structure of multivariate Poisson and its computational intractability \cite{MVPcomputing} non parametric 
mixture modeling for multivariate count vectors has received less attention.
Hence, we first bridge this gap by paralleling the development of \textit{DP based non-parametric 
mixture models and their temporal extensions  for multivariate count data} along the lines of those for Gaussians and multinomials .

Modeling the full covariance structure using the MVP is often computationally expensive. Hence, we further exploit 
the sparsity in data and introduce \textit{sparse mixture models for count vectors} and their temporal extensions.
We propose a sparse MVP Mixture 
modeling the covariance structure over a select subset of dimensions for each cluster. 
We are not aware of any prior work that models sparsity in count vectors. 

The proposed predictive models showed dramatic hitrate improvement on several real world traces. 
At the same time,  these count modeling techniques are of independent interest outside the caching problem 
as they can apply to a wide variety of settings such as text mining, where often counts are used.

{\bf Contributions:} 
Our first contribution, is the DP based non-parametric 
mixture of Multivariate Poisson (DP-MMVP) and its temporal extensions (HMM-DP-MMVP) which 
capture the full covariance structure of correlated count vectors. 
Our next contribution, is to exploit the sparsity in data, and proposing a novel technique for non parametric clustering 
 of sparse high dimensional count vectors with the sparse DP-mixture of Multivariate Poisson (Sparse-DP-MMVP).
This methodology not only leads to a better fit for sparse multidimensional count data but is also computationally more 
tractable than modeling full covariance. We then discuss a temporal extension, Sparse-HMM-DP-MMVP, 
for cache preloading. We are not aware of any prior work that addresses non-parametric modeling of sparse correlated count vectors. 

As our final contribution, we take the first steps in outlining a framework for cache preloading 
to capture long range spatio-temporal dependencies in memory accesses. 
We perform experiments on real-world benchmark traces showing dramatic hitrate improvements.
In particular, for the trace in Fig ~\ref{fig:intropica}, our preloading yielded 
a $0.498$ hitrate over $0.001$ of baseline (without preloading), a $498$X improvement 
(trace MT2: Tab \ref{tab:100feature}).
%

\section{Related Work}
\label{sec:related}

{\bf BNP for sparse correlated count vectors:}
\label{countdata}
Poisson distribution is a natural prior for count data. But the multivariate Poisson(MVP) \cite{MVPstructure} 
has seen limited use due to its computational intractability due to the complicated form of the 
joint probability function\cite{MVPcomputing}. There has been relatively little work on MVP mixtures \cite{MMVP,MVPmixture,MVPold, FishMVP}.  
On the important problem of designing MVP mixtures with an unknown number of components, \cite{MVPold} is the only reference  we are aware of.  
The authors explore MVP mixture with an unknown number of components using a truncated Poisson prior 
for the number of mixture components, and perform uncollapsed RJMCMC inference.
 DP based models are a natural truncation free alternative that are well studied and amenable to 
hierarchical extensions \cite{HDP,StickyHDPHMM} for temporal modeling which are of immediate 
interest to the caching problem. 
To the best of our knowledge, there has been no work that examines a truncation 
free non-parametric approach with  DP-based mixture modeling for MVP.
Another modeling aspect we address is the sparsity of data.  A full multivariate emission density over all 
components for each cluster may result in over-fitting due to excessive number of parameters introduced 
from unused components. We are not aware of any work on sparse MVP mixtures. 
There has been some work on sparse mixture models for Gaussian and multinomial densities \cite{sparseold} \cite{SparseTM}. 
However they are specialized to the individual distributions and do not apply here.
Finally, we investigate sparse MVP models for temporally correlated count vectors. We are not aware of 
any prior work that investigates mixture models for temporally correlated count vectors.

{\bf Cache Preloading:} 
Preloading has been studied before \cite{Bonfire} in the context of improving cache performance on enterprise storage servers 
for the problem of cache warm-up, of a cold cache by preloading the most recently accessed data, by analyzing block I/O traces. 
Our goal is however different, and more general, in that we are seeking to improve the cumulative hit rate by exploiting 
long ranging temporal dependencies even in the case of an already warmed up cache.
They also serve as an excellent reference for state of the art caching related studies 
and present a detailed  study of the properties of MSR traces.  We have used the same MSR traces as our benchmark. 
Fine-grained prefetching tehniques to exploit short range correlations \cite{quickminer,cminer}, some 
 specialized for sequential workload types \cite{SARC} (SARC) have been investigated in the past. 
 Our focus, however is to work with general non-sequential workloads, to capture long range access patterns, 
 exploring prediction at larger timescales.
Improving cache performance by predicting future accesses based on modeling file-system events was studied in \cite{ucsz}.
They operate over NFS traces containing details of file system level events. 
This technique is not amenable for our setting, where the only data source is block I/O traces, with no file system level data. 
\section{A framework for Cache Preloading based on mining Block I/O traces}
\label{sec:frame}
In this section we briefly describe the caching problem and describe our framework for cache preloading. 

\subsection{ The Caching Problem: } 
Application data is usually stored on a slower persistent storage medium like hard disk. 
A subset of this data 
is usually stored on {\bf cache}, a faster storage medium. 
When an application makes an I/O request for a specific block, if the requested block is in cache, 
it is serviced from cache. This constitutes a {\bf cache hit} with a low application latency  (in 
microseconds). Else, in the event of a {\bf Cache miss}, the requested block is first retrieved 
from hard disk into cache and then serviced from cache leading to much higher application latency
(in milliseconds).

Thus, the application's performance improvement is measured by $hitrate=\frac{\# cache hits}{\# cache hits + \# cache misses}$.
\subsection{The Cache Preloading Strategy: }
Our strategy involves observing a part of the trace $\mathcal D^{lr}$ for some period of time and deriving a model, 
which we term as {\bf Learning Phase}. We then use this model to keep predicting appropriate data to place in cache 
to improve hitrates in {\bf Operating Phase} at the end of each time slice ($\nu$ secs) for the rest of the trace  $\mathcal D^{op}$. 

In terms of the \textit{ execution time of our algorithms,} while the learning phase can take a few hours, 
the operational phase, is designed to run in time much less than the slice length of $\nu$ secs.
In this paper we restrict ourselves to learning from a fixed initial portion of the trace. 
In practice, the learning phase can be repeated periodically, or even done on an online fashion. 

{\bf Data Aggregation:}
As the goal is to improve hitrates by preloading data exploiting \textit{long range} dependencies, 
we capture this by \textit{aggregating} trace data into count vector sequences. 
We consider a partitioning of addressable memory (LBA Range) into $M$ equal bins. 
In the learning phase, we divide the trace  $\mathcal D^{lr}$ into $T_{lr}$ fixed length time interval slices of length $\nu$ seconds each.
Let $A_1, \hdots , A_{T_{lr}}$ be the set of actual access requests in each interval of $\nu$ seconds. 
We now aggregate the trace into a sequence of $T_{lr}$ count vectors $X_1, \hdots X_{T_{lr}} \in \mz^M$,  
each of M dimensions. Each count vector $X_t$ is a histogram of accesses in $A_t$ over M memory bins
in the $t$th time slice of the trace spanning $\nu$ seconds. 
%


\subsection{ Learning Phase (Learning a latent variable model):}
\label{modelingprob}
The input to the learning phase is a set of sparse count vectors $X_1, \hdots ,X_{T_{lr}} \in \mz^M$, correlated within 
and across instances obtained from a block I/O trace as described earlier.
These count vectors can often be intrinsically 
grouped into cohesive clusters which arise as a result of long range access patterns 
(see Figure \ref{fig:intropica}) that repeat over time albeit 
with some randomness. 
Hence we would like to explore unsupervised learning techniques based on clustering for these count vectors, 
that capture temporal dependencies between count vector instances and the correlation within instances.

Hidden Markov Models(HMM), are a natural choice of predictive models for such temporal data. In a HMM, latent variables 
$Z_t \in \{1,\hdots, K\}$ are introduced that follow a markov chain. 
Each $X_t$ is generated based on the choice of $Z_t$, 
inducing a clustering of count vectors.
In the learning phase, we learn the HMM parameters, denoted by $\theta$.

Owing to the variability of access patterns in trace data, a fixed value of K 
is not suitable for use in realistic scenarios motivating the use of non-parametric techniques of clustering. 
In section \ref{sec:model}  we propose the HMM-DP-MMVP, a temporal model for non-parametric 
clustering of correlated count vector sequences capturing their full covariance structure, followed by 
the Sparse-HMM-DP-MMVP in section \ref{SparseMVPModels}, that exploits the sparsity in count vectors to better model the data, also 
leading to more efficient inference.

As an outcome of the learning phase, we have a HMM based model with 
appropriate parameters, that provides predictive ability to infer the next hidden state 
on observing a sequence of count vectors. 
However, since the final prediction required is that of memory accesses, we maintain a map from 
every value of hidden state $k$ to the set of all raw access requests from various time slices during training 
that were assigned latent state k.
$H(k)= \cup_{\{t| Z_t =k\}} A_t$, for $\forall k$. 
\subsection{The Operating Phase (Prediction for Preloading): } 
Having observed $\{X_1, \ldots X_{T_{lr}}\}$ aggregated from  $\mathcal D^{lr}$, the learning phase learns a 
latent variable model.  In the Operating Phase, as we keep observing $\mathcal D^{op}$, after 
the time interval $t'$, the data is incrementally aggregated into a sequence  $\{X'_1, \hdots X'_{t'}\}$. 
At this point, we would like the model to predict the best possible 
choice of blocks to load into cache for interval $t'+1$ with knowledge of aggregated data $\{X'_1, \hdots X'_{t'}\}$. 

This prediction happens in two steps. In the first step, our HMM based model Sparse-HMM-DP-MMVP
infers hidden state $Z'_{t'+1}$ from observations $\{X'_1,\ldots,X'_{t'}\}$, 
using a Viterbi style algorithm as follows.
{\small
\begin{equation}
\label{prediction_problem}
(X'_{t'+1} ,\{Z'_{r}\}_{r=1}^{t'+1}) = \underset{(X'_{t'+1} ,\{Z'_{r}\}_{r=1}^{t'+1})}{\argmax} p(\{X'_{r}\}_{r=1}^{t'+1}, \{Z'_{r}\}_{r=1}^{t'+1} | \theta )   
\end{equation}
}
Note the slight deviation from usual Viterbi method as $X'_{t'+1}$ is not yet observed.
We also note that alternate strategies based on MCMC might be possible based on Bayesian techniques to infer the hidden 
state $Z'_{t'+1}$. However, \textit{in the operating phase, the execution time becomes important} and is required to be 
much smaller than $\nu$, the slice length. Hence we explore such a Viterbi based technique, that is quite efficient and 
runs in a very small fraction of $\nu$ in practice for each prediction.
The algorithm is detailed in the supplementary material. 

In the second step, having predicted the hidden state $Z'_{t'+1}$, we would now like to load the appropriate accesses. 
Our prediction scheme consists of loading all accesses defined by $H(Z'_{t'+1})$ into the cache (with H as defined previously).
%

\section{Mixture Models with Multivariate Poisson for correlated count vector sequences}
\label{sec:model}
We now describe non-parametric temporal models for correlated count vectors
based on the MVP \cite{MVPstructure} mixtures. 
MVP~\cite{MVPstructure} distributions are natural models for understanding multi-dimensional count data.
There has been no work on exploring DP-based mixture models for count data or for
modeling their temporal dependencies. 

Hence, we first parallel the development of non-parametric 
MVP mixtures along the lines of DP based multinomial mixtures 
\cite{HDP}. To this end we propose  DP-MMVP, a DP based MVP mixture and
propose a temporal extension HMM-DP-MMVP along the lines of HDP-HMM\cite{HDP} for multinomial mixtures. 

However a more interesting challenge lies in designing algorithms of scalable complexity for high dimensional
correlated count vectors. We address this in our next section( \ref{SparseMVPModels}) by introducing the Sparse-MVP that exploits sparsity in data.
 DP mixtures of MVP and their sparse counterparts lead to different inference challenges addressed
in section \ref{inference}.
\subsection{Preliminaries: } 
We first recall some definitions. 
A probability distribution $G \sim DP(\alpha, H)$, when 
$G = \sum_{k=1}^{\infty}  \beta_k \delta_{\theta_k} ,  \beta \sim GEM(\alpha),\; \theta_k \sim H, k=1 \hdots$
where $H$ is a diffused measure. 
 Probability measures $G_1$ $,\ldots, G_J$ follow {\bf Hierarchical Dirichlet process}(HDP)\cite{HDP} if   
\[ G_j \sim DP(\alpha , G_0) , j=1 \hdots J \mbox{\hspace{3pt} where \hspace{3pt}} G_0 \sim DP(\alpha,H)\]

HMMs are popular models for temporal data. However, for most applications there are no clear guidelines for fixing the number 
of HMM states. A DP based HMM model, {\bf HDP-HMM}, \cite{HDP} alleviats this need. Let $X_1, \hdots, X_T$ be observed data instances. 
Further, for any $L \in Z$, we introduce notation $[L]= \{1,\ldots,L\}$. The HDP-HMM is defined as follows. 
{ \centering $\beta \sim  GEM(\gamma) $}
\begin{gather*}
  {\bf \pi_k} | \beta, \alpha_k \sim DP(\alpha_k, \beta) ,\text{ and } \theta_k | H \sim H, k=1,2,\hdots \nonumber \\
  Z_t | Z_t-1,  \pi \sim  \pi_{Z_{t-1}}, \text{ and } X_{t} | Z_t \sim f_{Z_t}(\theta_k) , t \in [T]
\end{gather*}

Commonly used base distributions for  H  are the multivariate Gaussian and multinomial distributions. There has been no work in 
exploring correlated count vector emissions. In our setting, we explore MVP emissions with H being an appropriate 
prior for parameter $\theta_k$ of the MVP distribution.

{\bf The Multivariate Poisson(MVP):}
Let $\bar{a},\bar{b} > 0$.
A  random vector, $X \in \mz^M$  is Multivariate Poisson(MVP) distributed, denoted by~$X \sim MVP(\Lambda)$, if 
{\small
 \begin{align}
 \label{mvp}
 X = Y1_M \text{ Alternately, } X_j=\sum_{l=1}^M Y_{jl} , \forall j \in [M] \nonumber  \\
 {~where ~} \forall j \le l \in [M], \lambda_{l,j}=\lambda_{j,l} \sim Gamma(\bar a,\bar b) \nonumber \\
 Y_{j,l}= Y_{l,j}\sim Poisson(\lambda_{j,l})
 \end{align}
 }
and $1_M$ is a $M$ dimensional vector of all 1s. 
It is useful to note that $E(X)=\Lambda 1_M$, where $\Lambda$ is an $M \times M$ symmetric matrix with entries $\lambda_{j,l}$~ 
 and $Cov(X_{j},X_{l})=\lambda_{j,l}$.
Setting $\lambda_{j,l}=0 , j \neq l$ yields $X_i = Y_{i,i}$  
which we refer to as the {\bf Independent Poisson (IP)} model as $Y_{i,i}$ for each dimension $i$ are 
independently Poisson distributed. 
\subsection{DP Mixture of Multivariate Poisson (DP-MMVP): }
In this section we define DP-MMVP, a DP based non-parametric mixture model for clustering correlated count vectors. 
We propose to use a DP based prior,  $G \sim DP(\alpha, H)$, where $H$ is a suitably chosen Gamma 
conjugate prior for the parameters of MVP, $\Lambda=\{\Lambda_k : k=1,\hdots\} $, k being cluster identifier.
We define DP-MMVP as follows.
{\small
\begin{eqnarray}\label{eq:dmmvp}
\lambda_{kjl} \sim Gamma(\bar a,\bar b) , \forall  j \leq l \in [M], k=1, \hdots \nonumber \\
\beta \sim GEM(\alpha) \text{ and } G = \sum_{k=1}^{\infty}  \beta_k \delta_{\Lambda_k} \nonumber \\
Z_t |  \beta \sim Mult(\beta) \forall t \in [T] \nonumber \\
X_t | Z_t \sim MVP(\Lambda_{Z_t}), \forall t \in [T] 
\end{eqnarray}
}
where $T$ is the number of observations and $(\Lambda_k)_{jl}=(\Lambda_k)_{lj} = \lambda_{kjl}$.
%
%
%
%
%
We also note that the {\bf DP Mixture of Independent Poisson (DP-MIP)} can be similarly defined  by restricting 
$\lambda_{k,j,l}=0 , \forall j \neq l , k=1,\hdots$. 
\subsection{Temporal DP Mixture of MVP (HMM-DP-MMVP): }
\label{HMM-DP-MMVP}
DP-MMVP model does not capture temporal correlations that are useful for prediction problem of cache preloading. 
To this end  we propose HMM-DP-MMVP, a temporal extension of the previous model, as follows.
Let $X_t \in \mz^M, t \in [T]$ be  a temporal sequence of correlated count vectors.
{\small
\begin{gather}
\label{HMM-DP-MMVP}
\lambda_{kjl} \sim Gamma(\bar a,\bar b) \text{ } \forall j \leq l \in [M], k=1, \hdots \nonumber \\
\beta \sim GEM(\gamma) \hspace{3mm} {\bf \pi_k} | \beta, \alpha_k \sim DP(\alpha_k, \beta) \forall k= 1, \hdots \nonumber \\
Z_t | Z_t-1,  \pi \;\sim\;  \pi_{Z_{t-1}} ,  \forall t \in [T] \nonumber \\
X_t | Z_t \sim MVP(\Lambda_{Z_t}), \forall t \in [T] 
\end{gather}
}
The HMM-DP-MMVP incorporates the HDP-HMM structure into DP-MMVP in equation \eqref{eq:dmmvp}.
This model can again be restricted to the special case of diagonal covariance MVP
giving rise to the HMM-DP-MIP by extending the DP-MIP model. The HMM-DP-MIP models the temporal dependence, but not the spatial correlation
coming from trace data.
\section{Modeling with Sparse Multivariate Poisson: }
\label{SparseMVPModels}
We now introduce the \textit{Sparse Multivariate Poisson (SMVP)}.
Full covariance MVP, defined with ${M \choose 2}$ latent variables (in Y)
is computationally expensive during inference for higher dimensions.
However, vectors $X_t$ emanating from  traces are often very 
sparse with only a few significant components and most components 
close to $0$.  While there has been work on sparse multinomial\cite{STM} and sparse Gaussian\cite{SGM} mixtures, 
there has been no work on sparse MVP Mixtures. 
We propose the SMVP by extending the MVP to model sparse count vectors.
We then extend this to Sparse-DP-MMVP for a non-parametric mixture setting and finally propose  
the temporal extension Sparse-HMM-DP-MMVP.

\subsection{ Sparse Multivariate Poisson distribution (SMVP):}
We introduce the SMVP as follows. Consider an indicator vector $b \in \{0,1\}^M$,  that denotes whether a dimension is active or not.
Let $\hat{\lambda}_j \ge 0, \forall j \in [M]$ ~and~$b \in \{0,1\}^M$.   
We define $X \sim SMVP(\Lambda, {\bf \hat \lambda}, b)$ as: $X = Y1_M$ where $\forall j \leq l \in [M]$
{\small
\begin{equation}
Y_{j,l} \sim Poisson(\lambda_{j,l}) b_{j} b_{l} + Poisson(\hat \lambda_j) (1-b_{j}) \delta(j,l)) 
\end{equation}
}
where  $\Lambda$ is a symmetric positive matrix with $(\Lambda)_{jl} = \lambda_{jl}$.
If $b_j = 1, b_l =1$ then $Y_{j,l}$ is distributed as $Poisson(\lambda_{j,l})$.
However if $b_j = 0$, variables ~$Y_{j,j}$~are distributed as 
$Poisson(\hat{\lambda_j})$. The selection variables $b_j$ decide if the $j$th 
dimension is active. Otherwise we consider any emission at the $j$th dimension noise, 
modulated by $Poisson(\hat{\lambda}_j)$, independent of other dimensions. 
Parameter $\hat{\lambda_j}$ is close to zero for the extraneous noise dimensions and is common across clusters. 

With Sparse-MVP, we are defining a full covariance MVP for a subset of dimensions while 
the rest of the dimensions are inactive and hence modeled independantly (with a small mean to account for noise).
The full covariance MVP is a special case of Sparse-MVP where all dimensions are active.
%
%
%
\subsection{DP Mixture of Sparse Multivariate Poisson: }
In this section, we propose the {\bf Sparse-DP-MMVP}, extending our DP-MMVP model. 
For every mixture component k we introduce an indicator vector $b_k\in \{0,1\}^M$. 
Hence, $b_{k,j}$ denotes whether a dimension j is active for mixture component k.

A natural prior for selection variables, $b_{kj}$, is Bernoulli Distribution, while a Beta distribution is 
a natural conjugate prior for the parameter $\eta_j$ of the Bernaulli.
$ \eta_j \sim Beta(a', b'), b_{kj} \sim Bernoulli(\eta_j), j \in [M], k=1, \hdots. $ 
The priors for parameters $\Lambda$ and $\hat{\lambda}$ are again decided 
based on conjugacy, where $ \forall j \leq l \in [M]\; \lambda_{j,l}$ have a gamma prior, 
Let $\hat a, \hat b >0$.  We model ${\hat \lambda}$ to have 
a common Gamma prior for inactive dimensions over all clusters.
$\hat \lambda_j \sim Gamma(\hat a,\hat b), \forall j \in [M] $ .
The Sparse-DP-MMVP is defined as: 
{\small
\begin{gather}
\eta_j \sim  Beta(a',b'), \hspace{2mm} b_{k,j} \sim Bernoulli(\eta_j), j \in [M], k= 1 \hdots   \nonumber \\
  \hat \lambda_j \sim Gamma(\hat a, \hat b), j \in [M] \nonumber \\
\lambda_{kjl} \sim Gamma(\bar a, \bar b), \{ j \leq l \in [M] : b_{k,j}=b_{k,l}=1\}, k=1, \hdots  \nonumber \\
\vspace{5mm} 
G = \sum_{k=1}^{\infty}  \beta_k \delta_{\Lambda_k}, \beta \sim GEM(\alpha) \nonumber \\
\text{Cluster selection variables  } Z_t |  \beta \sim Mult(\beta) \text{  and} \nonumber \\
X_t | Z_t, \Lambda, {\bf \hat \lambda}, b_{Z_t}\sim SMVP(\Lambda_{Z_t},{\bf \hat \lambda},  b_{Z_t}) , t \in [T] 
\end{gather}
}

DP-MMVP is a special case of Sparse-DP-MMVP where all dimensions of all clusters are active.

\subsection{Temporal Sparse Multivariate Poisson Mixture: }
\begin{figure}[ht!]
\centering \includegraphics[height=1.9in,width=2.8in]{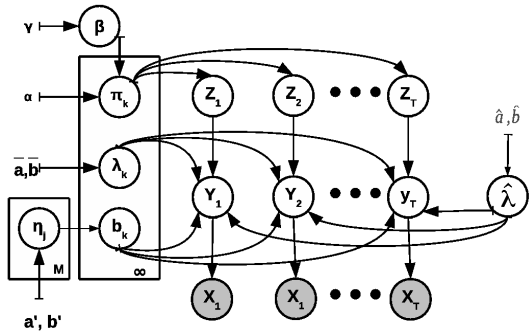}
\caption{\small \sl Plate Diagram : Sparse-HMM-DP-MMVP }  
\label{plate_sparse}
\end{figure}

We now define {\bf Sparse-HMM-DP-MMVP}, by extending Sparse-DP-MMVP to also capture
Temporal correlation between instances by incorporating HDP-HMM into the Sparse-DP-MMVP:
{\small
\begin{gather}
\label{EQ-Sparse-HMM-DP-MMVP}
  \eta_j \sim  Beta(a',b'), \hspace{2mm} b_{k,j} \sim Bernoulli(\eta_j), j \in [M], k= 1 \hdots   \nonumber \\
  \hat \lambda_j \sim Gamma(\hat a, \hat b), j \in [M] \nonumber \\
\lambda_{kjl} \sim Gamma(\bar a, \bar b), \{ j \leq l \in [M] : b_{k,j}=b_{k,l}=1\}, k=1, \hdots  \nonumber \\
\nonumber \\
\beta \sim GEM(\gamma) \text{ and}\hspace{2mm} {\bf \pi_k} | \beta, \alpha_k \sim DP(\alpha_k, \beta) , k= 1, \hdots  \nonumber \\
Z_t | Z_t-1,  \pi \sim  \pi_{Z_{t-1}} , t \in [T] \nonumber \\
X_{t} | Z_t, \Lambda_{Z_t}, {\bf \hat \lambda}, b_{Z_t}\sim SMVP(\Lambda_{Z_t},\hat \lambda,b_{Z_t}) , t \in [T]
\end{gather}
}
The plate diagram for the Sparse-HMM-DP-MMVP model is shown in figure \ref{plate_sparse}.
We again note that HMM-DP-MMVP model described in section \ref{HMM-DP-MMVP} is a restricted form of  Sparse-HMM-DP-MMVP where 
$b_{k,j}$ is fixed to 1. The Sparse-HMM-DP-MMVP captures the spatial correlation inherent in the trace data
and the long range temporal dependencies, at the same time exploiting sparseness, reducing the number of latent variables.

\section{Inference}
\label{inference}
While inference for non-parametric HMMs is well explored \cite{HDP}\cite{StickyHDPHMM}, 
MVP and Sparse-MVP emissions introduce additional challenges due to the latent variables involved in the definition 
of the MVP and the introduction of sparsity in a DP mixture setting for the MVP. 

We discuss the inference of  DP-MMVP in detail in the supplementary material. In this section, we give a brief 
overview of collapsed Gibbs sampling inference for HMM-DP-MMVP and Sparse-HMM-DP-MMVP. More details are again 
in the supplementary material.

Throughout this section, we use the following notation: $Y=\{Y_t : t \in [T]\}$, $Z=\{Z_t : t \in [T]$ and $X=\{X_t : t \in [T]$.
A set with a subscript starting with a hyphen($-$) indicates the  set of all elements except the index following the hyphen.
The latent variables to be sampled are  
$Z_{t}, t \in [T]$, 
$Y_{t,j,l}, j \leq l \in[M], t \in [T]$ and $b_{k,j} ,  j \in [M], k=1, \hdots $ (For the sparse model).
Further we have $\beta=\{\beta_1, \hdots, \beta_K, \beta_{K+1}= \sum_{r=K+1}^\infty \beta_r\}$. 
and an auxiliary variable $m_{k}$ is introduced as a latent variable to aid the sampling of $\beta$ based on the direct sampling procedure from HDP\cite{HDP}. 
Updates for  $m$ and $\beta$ are similar to \cite{StickyHDPHMM}, as detailed in algorithm \ref{alg:inference}.
The latent variables $\Lambda$ and  $\pi$ are collapsed.

\underline{Sampling $Z_{t}, t \in [T]$}
While the update for this variable is similar to that in \cite{HDP}\cite{StickyHDPHMM}, the likelihood term 
$p(Y | Z_t=k,Z_{-t}, Y_{-t}, \bf b^{old}; \bar a, \bar b)$ differs due to the MVP based emissions. 
For the HMM-DP-MMVP, this term can be evaluated by integrating out the $\lambda$s. Similarly for Sparse-HMM-DP-MMVP
when $k$ is an existing componant (for which $b_k$ is known).

However, for the Sparse-HMM-DP-MMVP, an additional complication arises for the case of a new componant, 
since we do not know the $b_{K+1}$ value, requiring summing over all possibilities of $b_{K+1}$, leading to exponential complexity.
Hence, we evaluate this numerically (see supplementary material for details). This process is summarized in algorithm \ref{alg:inference}.

\underline{Sampling $Y_{t,j,l}, j \leq l \in[M], t \in [T]$} : 
The  $Y_{t,j,l}$ latent variables in the MVP definition, differentiate the 
inference procedure of an MVP mixture from standard inference for DP mixtures. Further, the large number of $Y_{t,j,l}$
variables $n \choose{2}$ also leads to computationally expensive inference for higher dimensions motivating sparse modeling.
In, Sparse-HMM-DP-MMVP, only those  $Y_{t,j,l}$ values are updated for which $b_{Z_t,j}=b_{Z_t,l}=1$.

We have, for each dimension j, $X_{t,j} = \sum_{l=1}^M Y_{t,j,l}$. To preserve this constraint, 
suppose for row j, we sample ${Y_{t,j,l}, j \neq l}$, $Y_{t,j,j}$ becomes a derived quantity as
$Y_{t,j,j} = X_{t,j} - \sum_{p=1, p \neq j}^{M} Y_{t,p,j}  $.
We also note that, updating the value of $Y_{t,j,l}$ impacts the value of only two other random variables
i.e $Y_{t,j,j}$ and $Y_{t,l,l}$.
The final update for $Y_{t,j,l}, j \neq l$ can be obtained by integrating out $\Lambda$. 
(full expression in alg \ref{alg:inference}, more details: Appendix B).

\setlength{\algomargin}{-1pt}
\begin{algorithm}[!ht]
{\footnotesize
\text{{\bf Repeat} until convergence}\\
\For {$t=1, \hdots , T$} {
  \underline{// Sample $Z_t$} from
\begin{gather}
 \label{inf:sparse:samplez}
 p(Z_t =k | Z_{-t,-(t+1)},z_{t+1}=l, {\bf b^{old}}, X, \beta , Y; \alpha, \bar a, \bar b)  \nonumber  \\
\propto p(Z_t =k, Z_{-t,-(t+1)},_{t+1}=l | \beta ; \alpha, \bar a, \bar b) \nonumber \\
 p(Y | Z_t=k,Z_{-t}, Y_{-t}, \bf b^{old}; \bar a, \bar b)   \nonumber
 \end{gather}
 \text{//Case 1: For HMM-DP-MMVP (with $b_{k,j}=1 \text{ } \forall j,\forall k$) }
 \text{//and Sparse-HMM-DP-MMVP For existing k}
\begin{gather} \label{Sparse_yall_update}
p(Y | Z_t=k,Z_{-t}, Y_{-t}, {\bf b^{old}}; \bar a, \bar b) \propto \underset {  j \leq l \in [M]} {\Pi} F_{k,j,l} ^{b_{k,j} b_{k,l}} \underset { j \in [M]} {\Pi} \hat F_{j} \nonumber \\
\label{hatfkjl}
\text{Where } F_{k,j,l}= \frac{\Gamma(\bar a+S_{k,j,l})}{(\bar b+n_k)^{(\bar a+S_{k,j,l})} \underset{\bar t: Z_{\bar t}=k} {\Pi} Y_{\bar t,j,l}!} \nonumber \\
\text{and } \hat F_{j}= \frac{\Gamma(\hat a+\hat S_{j})}{(\hat b+\hat n_j)^{(\hat a+\hat S_{j})} \underset{t,j: b_{Z_t,j}=0} {\Pi} Y_{t,j,j}!} \nonumber
\end{gather},
With $\hat S_{j}=\sum_t  Y_{t,j,j} (1-b_{Z_t,j})$ , $\hat n_{j}=\sum_t  (1-b_{Z_t,j})$  and
$S_{k,j,l}=\sum_{\bar t} Y_{\bar t,j,l} \delta(Z_{\bar t},k)$, for $j \leq l \in [M] $ 
 \text{//Case 2: For Sparse-HMM-DP-MMVP for new k, }
 \text{//compute following numerically,where ${\bf b^{old}}=\{b_1,\hdots,b_K\}$}
\begin{align*}
p(Y | {\bf b^{old}}, Z_t=K+1,Z_{-t}, Y_{-t}; \bar a, \bar b) = & \\
\sum_{b_{K+1}} p(b_{K+1} | {\bf b^{old}}, \eta) \\
p(Y |{\bf  b^{old}}, b_{K+1},  Z_t=K+1,Z_{-t}, Y_{-t}; \bar a, \bar b)
\end{align*}

  \For{$ j \leq l \in [M] $}{
     \If{$b_{Z_t,j}=b_{Z_t,k}=1$}{
         \underline{// Sample $Y_{t,j,l}$} from 
         \begin{align}
p(Y_{t,j,l} | Y_{-t,j,l}, Z, \bar a, \bar b) \propto 
 F_{k,j,l} F_{k,j,j} F_{k,l,l}   \nonumber 
\end{align}
         Set $Y_{t,j,j}=X_{t,j}-\sum_{\bar j=1}^M Y_{t,j,\bar j}$ \\
         Set $Y_{t,l,l}=X_{t,l}-\sum_{\bar l=1}^M Y_{t,l,\bar l}$ 
     }
  }  
}
\For{$j=1, \hdots, M, k=1, \hdots, K$}{
  \underline{// Sample $b_{k,j}$ } (for Sparse-HMM-DP-MMVP) from 
  \begin{equation*}
p(b_{k,j} | b_{-k,j}, Y, Z ) \sim p(b_{k,j} | b_{-k,j}) p(Y| b_{k,j}, b_{-k,j}, Z ; \bar a \bar b)
\end{equation*}
  \[p(b_{k,j} | b_{-k,j}) \propto \frac{c_j^{-k} + b_{k,j} + a' -1}{K + a' + b' -1}\] 
  where $c_j^{-k}=\sum_{\bar k \neq k, \bar k=1}^K b_{k,j}$ is the  number of clusters (excluding k) with dimension $j$ active
}
 \For{$k=K, \hdots, M, k=1, \hdots$}{
   $m_k=0$ \\
   \For {$i=1, \hdots , n_k$} {
   $u \sim Ber(\frac{\alpha \beta_k}{i+\alpha \beta_k})$, if $(u==1)m_k++$ \\
   }
 }
 $[\beta_1 \beta_2 \hdots \beta_K \beta_{K+1}]  | m, \gamma \sim Dir(m_1, \hdots , m_k, \gamma)$
 
 \caption{\scriptsize Inference: Sparse-HMM-DP-MMVP \newline
  Inference steps(The steps for HMM-DP-MMVP are similar and \newline
  are shown as  alternate updates in brackets)}
  \label{alg:inference}
 }
\end{algorithm}

\underline{\textbf{ Update for $b_{k,j}$:}}
For Sparse-HMM-DP-MMVP, the update for $b_{k,j}$ is obtained by integrating out $\eta$ to evaluate $p(b_{k,j} | b_{-k,j})$
and computing the likelihood term by integrating out $\lambda, \hat \lambda$ as before. This is shown in algorithm \ref{alg:inference} (see supplementary
material for more details).

\textbf{Train time Complexity Comparison: Sparse-HMM-DP-MMVP vs HMM-DP-MMVP: }
The inference procedure for both models is similar, with different updates shown in Algorithm \ref{alg:inference}. For the HMM-DP-MMVP all 
dimensions are active for all clusters. We sample  $M \choose 2$  random variables for  the symmetric matrix $Y_t$ in this step for 
each $t \in [T]$. On the other hand, for Sparse-HMM-DP-MMVP with $\bar m_k$ active components in cluster k, we sample only $\bar m \choose 2$
random variables which is a significant improvement when $\bar m << M$.

\section{Experimental Evaluation}
\label{experiments}

\begin{figure*}[!Htbp]
\centering
\includegraphics[height=1.5in,width=6.7in]{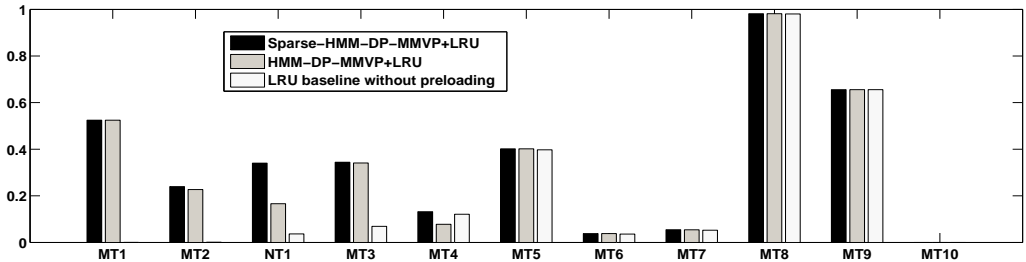}
\caption{\small \sl Comparison of Hitrates against LRU baseline without preloading : 
On MT1 we show the highest hitrate increase of 0.52 vs baseline 0.0002. On MT10 we show the least improvement
where both baseline and our model yield 0 hitrate.}  
\label{hitrates}
\vspace{-1em}
\end{figure*}  
We perform experiments on benchmark traces, to evaluate our models, 
in terms of likelihood and also evaluate their effectiveness for the caching problem, by 
 measuring hitrates using our predictive models.

\subsection{Datasets: }

We perform experiments on diverse enterprise workloads : 10 publicly available real world Block I/O traces (MT 1-10),
commonly used benchmark in storage domain, 
collected at Microsoft Research Cambridge\cite{MSFTdataset}
and 1 NetApp internal workload (NT1).
See dataset details and choice of traces in Appendix C.

We divide the available trace into two parts $\mathcal D^{lr}$ that is aggregated into $T_{lr}$ 
count vectors and $\mathcal D^{op}$ that is aggregated into $T_{op}$ count vectors.
In our initial experimentation, for aggregation, we fix the number of memory bins M=10 (leading to 10 dim count vectors), 
length of time slice $\nu$=30 seconds. Further, we use a test train split of 50\% for both experiments such that $T_{lr}=T_{op}$.
(Later, we also perform some experiments to study the impact of some of these parameters  with $M=100$ dimensions on some of the traces.)

\subsection{Experiments: } 
We perform two types of experiments, to understand how well our model fits data in terms of likelihood,
the next to show how our model and our framework can be used to improve cache hitrates.

\subsubsection{  Experiment Set 1: Likelihood Comparison: } 
We show a likelihood comparison between HMM-DP-MMVP, Sparse-DP-MMVP and baseline model HMM-DP-MIP.
We train the three models using the inference procedure detailed in section \ref{inference} 
on $\mt_{lr}$ ~and compute Log-likelihood on the held out test trace $\mt_{op}$. The results are tabulated in Table ~\ref{tab:likel}. 

{\bf Results:} We observe that the Sparse-HMM-DP-MMVP model performs the best in terms of likelihood, while the HMM-DP-MMVP outperforms 
HMM-DP-MIP by a large margin. Poor performance of HMM-DP-MIP clearly shows that spatial correlation present across the $M$ 
dimensions is an important aspect and validates the necessity for the use of a Multivariate Poisson model over an 
independence assumption between the dimensions. Superior performance of Sparse-HMM-DP-MMVP over HMM-DP-MMVP is again 
testimony to the fact that there exists inherent sparsity in the data and this is better modeled by the sparse model. 
\vspace{-1em}
\begin{table}[htbp]
\centering
\caption{Log-likelihood on held out data: 
Sparse-HMM-DP-MMVP model fits data with Best Likelihood }
\label{likelihood}
\resizebox{5.5cm}{!} {
\begin{tabular}{| l | c|  r| r|r|r|}\hline
Trace   & HMM-DP & HMM-DP  & Sparse-HMM \\
 Name             & IP & MMVP  & DP-MMVP \\
    &($\times 10^6$)&  ($\times 10^6$) &   ($\times 10^6$)  \\\hline
NT1&-24.43&  -20.18 &    {\bf -19.35}  \\
MT1  & -14.02  & -8.12  &{\bf -8.10 }   \\
MT2 & -0.46   & -0.44  & {\bf -0.27 }   \\
MT3 &-0.09 & -0.08  &    {\bf -0.06 }    \\
MT4 &-1.15   & -1.01   & {\bf -0.93 }     \\ 
MT5 &-20.23   & -12.70 & {\bf -12.61}    \\
MT6 &-69.12 &-50.86 &    {\bf -50.53 }   \\ 
MT7  &-87.25    &-84.97& {\bf -80.24 }    \\
MT8 &-2.44 & -2.33  &   {\bf -2.03  }    \\ 
MT9 &-12.45  & -12.85 & {\bf -11.53 }    \\  
MT10 &-16.49 & -13.52 &  {\bf -12.91 }   \\  \hline  
\end{tabular}
}
\label{tab:likel}
\end{table}
\subsubsection{ Experiment Set 2: Hitrates: }
We compute hitrate, on each of the 11 traces,  with a
baseline simulator without preloading and an augmented simulator with the ability to preload predicted blocks every 
$\nu =30s$. Both the simulators use LRU for eviction.
Off the shelf simulators for preloading are not available for our purpose and  construction of the baseline simulator and that 
with preloading are described in detail in supplementary material- Appendix C.

{\bf Results}: We see in the barchart in figure \ref{hitrates} prediction improves hitrates over LRU baseline without preloading. 
We see that our augmented simulator gives order of magnitude better hitrates on certain traces (0.52 with preloading against 0.0002 with plain LRU).

{\bf Effect of Training Data Size: } 
We expect to capture long range dependencies in access patterns when we observe a sufficient portion of 
the trace for training where such dependencies manifest. Ideally we would like to run our algorithm in an online setting, 
where periodically, all the data available is used to update the model. Our model can be easily adapted to such a situation. 
In this paper, however, we experiment in a setting where we always train the model on 50\% (see supplementary material for an explanation
of the figure 50\%) of available data for each trace
and use this model to make predictions for the rest of the trace.

{\bf Effect of M (aggregation granularity): } To understand the impact M (count vector dimensionality), we pick
some of the best performing traces from the previous experiment (barchart in figure \ref{hitrates}) and repeat our experiment with  M=100 
features, a finer 100 bin memory aggregation leading to 100 dimensional count vectors.
We find that we beat baseline by an even higher margin with M=100. 
We infer this could be attributed to the higher sensitivity of our algorithm to detail in traces leading to superior clustering. 
This experiment also brings to focus the \textit{training time} of our algorithm. We observed that the Sparse-HMM-DP-MMVP outperforms HMM-DP-MMVP not only in terms of likelihood and 
hitrates but also in terms of training time. We fixed the training time to at most 4 hours to run our algorithms and report hitrates in 
table \ref{tab:100feature}. We find that  Sparse-HMM-DP-MMVP ran to convergence while HMM-DP-MMVP did not finish even a single 
iteration for most traces in this experiment. This corroborates our understanding that handling sparsity reduces the number of latent variables  
in HMM-DP-MMVP, improving inference efficiency translating to faster training time, particularly for higher dimensional count vectors.
\vspace{-2em}
 \begin{table}[hbt]
 \centering
 \caption{Hitrate after training 4 hours: M=10,M=100(x indicates that
inference did not complete one iteration)}
\resizebox{8cm}{!} {
 \begin{tabular}{ | l |c| r|r|r|r|} \hline
  Trace        &S-HMM-&S-HMM-&HMM-&LRU\\ 
  Name         &DP-MMVP&DP-MVP&DP-MVP&without\\ 
               &M=10&M=100&M=100&Preloading\\ \hline
  NT1 &0.340   & 0.592 (43 clusters)& 0.34& 0.0366  \\
  MT1 &0.5245  & 0.710 (190 clusters)&x & 0.0002\\ 
  MT2 & 0.2397 & 0.498 (23 clusters)&x  & 0.0010\\ 
  MT3 &0.344   & 0.461 (53 clusters)&x & 0.0366\\  \hline
  Avg &0.362   & 0.565 &- &0.0186 \\ \hline
 \end{tabular}
 }
 \label{tab:100feature}
 \end{table}
\subsection{Discussion of Results: }
We observe both from table \ref{likelihood} and the barchart (fig \ref{hitrates}) that  HMM-DP-MMVP 
outperforms HMM-DP-MIP, and Sparse-HMM-DP-MMVP performs the best, outperforming HMM-DP-MMVP in terms of likelihood 
and hitrates, showing that traces indeed exhibit
spatial correlation that is effectively modeled by the full covariance MVP and that 
 handling sparsity leads to a better fit of the data.

The best results  are tabulated in  Table \ref{tab:100feature} where we observe that when using $100$ bins Sparse-HMM-DP-MVP 
model achieves an average hitrate of $h = 0.565$, 30 times improvement over LRU without preloading, $h = 0.0186$. 
On all the other traces, LRU without preloading is  outperformed by the sparse-HMM-DP-MMVP, the improvement 
being dramatic for 4 of the traces. On computing average hitrate for 
the 11 traces in figure \ref{hitrates}, we see 58\% hitrate improvement.

{\bf Choice of Baselines: } We did not consider a parametric baseline as it is clearly not suitable for our caching application. Traces have different access patterns with 
varying detail (leading to varying number of clusters: fig \ref{tab:100feature}). A parametric model is clearly infeasible in a realistic scenario. Further, due to lack of 
existing predictive models
for count vector sequences, we use a HMM-DP-MIP baseline for our models.

{\bf Extensions and Limitations: }While we focus on capturing long range correlations,  our framework can be augmented with 
other algorithms geared towards specific workload types for capturing short range correlations, like sequential 
read ahead and Sarc \cite{SARC} to get even higher hitrates. We hope to investigate this in future.

We have shown that our models lead to dramatic improvement for a subset of traces and work 
 well for the rest of our diverse set of traces. 
We note that there may not be discernable long range correlations present 
in all traces. However, we have shown, 
that when we can predict, the scale of its impact is huge. 
There are occasions, when the prediction set is larger than cache where we would have to understand the 
temporal order of predicted reads, to help efficiently schedule preloads. 
Cache size, prediction size and preload frequency, all play an important role to evaluate the 
full impact of our method. Incorporating our models  within a full-fledged storage system involves further challenges, 
such as real time trace capture, smart disk scheduling algorithms for preloading, etc.  
These issues are beyond the scope of the paper, but form the basis for future work.
\section{Conclusions}
\label{conclusion}
 We have proposed DP-based mixture models (DP-MMVP, HMM-DP-MMVP) for correlated count vectors that capture the full covariance structure of 
 multivariate count data. We have further explored the sparsity in our data and proposed models 
 (Sparse-DP-MMVP and Sparse-HMM-DP-MMVP) that capture the correlation within a subset of dimensions for each cluster,
 also leading to more efficient inference algorithms.
 We have taken the first steps in outlining a preloading framework for leveraging long range 
 dependencies in block I/O Traces to improve cache hitrates. Our algorithms achieve a 30X hitrate improvement on 
 $4$ real world traces, and outperform baselines on all traces. 
\vspace{-1em}
\bibliographystyle{abbrv}

{\scriptsize
\bibliography{clusteringhistograms}}
\pagebreak
\twocolumn[ 
\begin{@twocolumnfalse} 
  \title{\Large Supplementary Material: Mining Block I/O Traces for Cache Preloading with  Sparse Temporal Non-parametric Mixture of Multivariate Poisson}             
  \vspace{1em}
\end{@twocolumnfalse}
]

\section*{\fontsize{14}{15}\selectfont Appendix A: Prediction Method}
 \addtocounter{section}{1}
\label{sup:appendix1} 
The prediction problem in the operational phase involves finding the best $Z'_{t+1}$ using  $\theta$ to solve equation \ref{prediction_problem}. 
We describe a Viterbi like dynamic programming algorithm to solve this problem  for Sparse-MVP emissions.
(A similar procedure can be followed for MVP emissions).

We note that alternate strategies based on MCMC might be possible based on Bayesian inference for the variable under 
question. However, in the operation phase, the execution time becomes important and is required to be much smaller than $\nu$, 
the slice length. Hence we explore the following dynamic programming based procedure that is efficient and runs in a 
small fraction of slice length $\nu$.

At the end of the learning phase, we estimate the values of ${\bf \theta}=\{\Lambda, \pi, b, \hat \lambda\}$, by obtaining $\Lambda, \hat \lambda, \pi $
as the mean of their posterior, and $b$ by thresholding the mean of its posterior and use these as parameters during prediction. 
A standard approach to obtain the most likely decoding of the hidden state sequence 
is the Viterbi algorithm, a commonly used dynamic programming  technique that finds
\[{\{Z'^*_s\}}_{s=1}^{t} = \underset{{\{Z'_s\}}_{s=1}^{t}}{\argmax} \hspace{2mm} p(X'_1,... X'_{t'}, {\{Z'_s\}}_{s=1}^{t}) | {\bf \theta} )   \]

Let $\omega(t,k)$ be the highest probability along a single path ending with $Z'_t=k$. Further, let 
$\omega(t,k)=\underset{{\{Z'_s\}}_{s=1}^{t}}{\max} p({\{X'_s\}}_{s=1}^{t}, {\{Z'_s\}}_{s=1}^{t}) | {\bf \theta} ) $. 
We have 
\[\omega(t+1,k)= \underset{k'=1, \hdots, K}{\argmax} \hspace{1mm} \omega(t,k') \pi_{k',k} SMVP(X'_{t+1} ; \theta) \]

Hence, in the standard setting of viterbi algorithm, having observed $X'_{t+1}$, 
the highest probability estimate of the latent variables is found as 
${Z'}^*_{t+1}= \underset{1 \leq k \leq K}{\argmax}\hspace{1mm} \omega(t+1,k)$.
However, the evaluation of MVP and hence the evaluation of the SMVP pmf involves exponential complexity due to integrating out the Y variables. 
While there are dynamic programming based approaches explored for MVP evaluation \cite{MVPcomputing}, we resort to a simple 
 approximation. Let $\mu_{k,i}=\sum_{j=1}^M \lambda_{k,i,j} b_{k,i} b_{k,j}+(1-b_{k,i}) \hat \lambda_j$, $i \in [M]$, $k \in [K]$. We consider
$X_{t,i} | Z_t=k \sim Poisson(\mu_{k,i})$ when  $X_t \sim SMVP(\Lambda_k, \hat \lambda)$ (since the sum of independent Poisson random variables
is again a Poisson random variable). 
Hence we compute $p(X_t | Z_t=k, \mu_k)= \Pi_{i=1}^M Poisson(X_{t,i} ; \mu_{k,i})$.

In our setting, \textit {we require finding the most likely $Z'^*_{t+1}$ without having observed $X'_{t+1}$} to address our prediction problem from section
\ref{sec:frame}. Hence we define the following optimization problem that tries to maximize the objective function over 
the value of $X'_{t+1}$ along with the latent variables ${\{Z'_s\}}_{s=1}^{t+1}$.
\[\omega'(t+1,k) = \underset{{\{Z'_s\}}_{s=1}^{t+1},X'_{t+1}}{\max} p({\{X'_s\}}_{s=1}^{t+1}, {\{Z'_s\}}_{s=1}^{t+1}) | {\bf \theta} ) \]

However, since mode of Poisson is also its mean, 
\begin{equation}
\label{sup:dynprog}
\omega'(t+1,k) = Poisson(\mu_k | \mu_k) \underset{k=1, \hdots , K}{\max} \omega'(t,k) \pi_{k,l} 
\end{equation}

From equation \ref{sup:dynprog}, we have a dynamic programming algorithm similar to Viterbi algorithm (detailed in algorithm \ref{sup:viterbi}).


%
%
\setlength{\algomargin}{2pt}
\begin{algorithm}
\small
\bf{Initial Iteration: }Before $X_1$ is observed  \\
 $\omega'(1,k)=\pi^0_k  Poisson(\mu_k ; \mu_k) \forall k$ \\
 $Z^*_1 = Argmax_k \hspace{2mm} \omega'(1,k)$ \\
\bf{Initial Iteration: }After $X_1$ is observed  \\
 $\omega(1,k)=\pi^0_k Poisson(X_1 ; \mu_k) \forall k$  \\
 \For{ $t=2, \hdots T$  }{
   \bf{Before $X_t$ is observed} \\
   $\omega'(t,l)=\max_k (\omega(t-1,k) \pi_{kl}) Poisson(\mu_l, \mu_l) \forall k$ \\
   ${Z^*_t} = Argmax_k  \hspace{2mm} \omega'(t,l)$    \\
   \bf{After $X_t$ is observed} \\
   $\omega(t,l)=\max_k (\omega(t-1,k) \pi_{kl}) Poisson(X_t,\mu_l) \forall k$  \\
   $\Psi(t,l) = Argmax_k  \hspace{2mm} (\omega(t-1,k)  \pi_{kl})$ \\
 }
\bf{Finding the Path} 
 $Z_T=Argmax_k \omega(t,K)$  \\
\For{ Data points  $t=T-1, T-2 \hdots$ 1 }{
   $Z_T=\Psi(t+1,Z_(t+1))$ 
 }
 \caption{Prediction Algorithm}
 \label{sup:viterbi}
\end{algorithm}

\vspace{3em}
\section*{\fontsize{14}{15}\selectfont Appendix B: Inference Elaborated}
 \addtocounter{section}{1}
\label{sup:inference-full}
In this section of supplementary material we discuss the inference procedure for DP-MMVP, HMM-DP-MMVP and Sparse-HMM-DP-MMVP
more elaborately adding some details that could not be accomodated in the original paper.
Our Collapsed Gibbs Sampling inference procedure is described in the rest of this section.

We first outline the inference for DP-MMVP model in section \ref{sup:inference-DP-MMVP} , followed by the HMM-DP-MMVP, 
its temporal extension in section \ref{sup:inference-HMM-DP-MMVP}. Then, in section \ref{sup:inference_Sparse-HMM-DP-MMVP}, 
we describe the inference for the Sparse-HMM-DP-MMVP model extending the previous procedure.
\subsection{Inference : DP-MMVP: }
\label{sup:inference-DP-MMVP}
The existance of $Y_{t,j,l}$ latent variables in the MVP definition differentiates the 
inference procedure of an MVP mixture from standard inference for DP mixtures. (The large number of $Y_{t,j,l}$
variables also leads to computationally expensive inference for higher dimensions motivating sparse modeling).

We collapse $\Lambda$ variables exploiting the Poisson-Gamma conjugacy for faster mixing. 
The latent variables $Z_{t}, t \in [T]$, and $Y_{t,j,l}, j \leq l \in[M], t \in [T]$ require to be sampled.
Throughout this section, we use the following notation: $Y=\{Y_t : t \in [T]\}$, $Z=\{Z_t : t \in [T]$ and $X=\{X_t : t \in [T]$.
A set with a subscript starting with a hyphen($-$) indicates the  set of all elements except the index following the hyphen.

\underline{\textbf{ Update for $Z_t$:}}
\label{sup:inference_Z_DP-MMVP}
The update for cluster assignments $Z_t$ are based on the conditional obtained on integrating out G, based on the  CRP\cite{CRP} process
leading to the following product.
\begin{gather}
\label{sup:DP-MMVP-updatez}
p(Z_t =k | Z_{-t}, X, \beta , Y; \alpha, \bar a, \bar b)  \propto p(Z_t =k | Z_{-t} ; \alpha)  f_k(Y_t) \nonumber \\
\propto \begin{cases}
   n^{-t}_{k} f_k(Y_t) & \text{k $\in [K]$}  \\
   \alpha f_k(Y_t) & \text{k=K+1 }   
  \end{cases}
\end{gather}
Where $n_{k,}^{-t}=\sum_{\bar t \neq t} \delta(Z_{\bar t}, k) $.
The second term $f_k(Y_t)=p(Y_t, Y_{-t} | Z_t=k,Z_{-t}; \bar a, \bar b)$ can be simplified by integrating out the $\Lambda$ variables based on 
their conjugacy. \\
Let $S_{k,j,l}=\sum_{\bar t} Y_{\bar t,j,l} \delta(Z_{\bar t},k)$, for $j \leq l \in [M] $  and
\begin{gather} 
\label{sup:fkjl}
F_{k,j,l}= \frac{\Gamma(\bar a+S_{k,j,l})}{(\bar b+n_k)^{(\bar a+S_{k,j,l})} \underset{\bar t: Z_{\bar t}=k} {\Pi} Y_{\bar t,j,l}!} \\
\label{sup:zyterm}
\text{By collapsing $\Lambda$, }
f_k(Y_t) \propto
\underset { 1<=j<=l<=M} {\Pi} F_{k,j,l} 
\end{gather}

\underline{\textbf{ Update for $Y_{t,j,l}$:} }
\label{sup:inference_Y_DP-MMVP}
This is the most expensive step since we have to update ${M \choose 2}$ variables for each observation $t$. 
The $\Lambda$ variables are collapsed, owing to the Poisson-Gamma conjugacy
due to the choice of a gamma prior for the MVP.

In each row $j$ of $Y_t$, $X_{t,j} = \sum_{l=1}^M Y_{t,j,l}$. To preserve this constraint, 
suppose for row j, we sample ${Y_{t,j,l}, j \neq l}$, $Y_{t,j,j}$ becomes a derived quantity as
$Y_{t,j,j} = X_{t,j} - \sum_{p=1, p \neq j}^{M} Y_{t,p,j}  $.

The update for $Y_{t,j,l}, j \neq l$ can be obtained by integrating out $\Lambda$ to get an expression similar to that in 
equation \ref{sup:fkjl}. We however note that, updating the value of $Y_{t,j,l}$ impacts the value of only two other random variables
i.e $Y_{t,j,j}$ and $Y_{t,l,l}$. Hence we get the following update for $Y_{t,j,l}$
\begin{align}
\label{sup:yupdate}
p(Y_{t,j,l} | Y_{-t,j,l}, Z, \bar a, \bar b) \propto 
 F_{k,j,l} F_{k,j,j} F_{k,l,l}  
\end{align}

The support of $Y_{t,j,l}$, a positive, integer valued random variable,  can be restricted as follows for efficient computation.
We have $Y_{t,j,j}= X_{t,j} - \sum_{p=1, p \neq j}^{M} Y_{p,j}  \geq 0 $
Similarly,  $Y_{t,l,l}= X_{t,l} - \sum_{p=1, p \neq l}^{M} Y_{p,l} \geq 0$.
Hence, we can reduce the support of $Y_{t,j,l}$ to the following: 
\begin{equation}
\label{sup:support}
0 \leq Y_{t,j,l} \leq min\left(( X_{t,j}  - \sum_{p=1, p \neq l}^{M} Y_{p,j} ), 
( X_{t,l}  - \sum_{p=1, p \neq j}^{M} Y_{p,l} )\right)
\end{equation} 
\subsection{Inference : HMM-DP-MMVP: }
\label{sup:inference-HMM-DP-MMVP}
The latent variables from the HMM-DP-MMVP model that require to be sampled include 
$Z_{t}, t \in [T]$, 
$Y_{t,j,l}, j,l \in[M], t \in [T]$ ,
and $\beta=\{\beta_1, \hdots, \beta_K, \beta_{K+1}= \sum_{r=K+1}^\infty \beta_r\}$. 
Additionally an auxiliary variable $m_{k}$ (denoting the cardinality of the partitions generated by the base DP) is introduced as a latent variable 
to aid the sampling of $\beta$ based on the direct sampling procedure from HDP\cite{HDP}. The latent variables $\Lambda$ and  $\pi$ are collapsed to facilitate faster mixing.
The procedure for sampling of $Y_{t,j,l}, j,l \in[M], t \in [T]$ is the same as that for DP-MMVP (eq: \ref{sup:yupdate}).
Updates for  $m$ and $\beta$ are similar to \cite{StickyHDPHMM}, detailed in algorithm \ref{alg:inference}.
We now discuss the remaining updates.

\underline{\textbf{ Update for $Z_t$:}}
\label{sup:inference_Z_HMM-DP-MMVP}
The update for cluster assignment for the HMM-DP-MMVP while similar to that that of DP-MMVP also considers the  
temporal dependency between the hidden states. Similar to the procedure outlined in \cite{HDP}\cite{StickyHDPHMM} we have:
\begin{gather}
\label{sup:updatez}
p(Z_t =k | Z_{-t,-(t+1)},z_{t+1}=l, X, \beta , Y; \alpha, \bar a, \bar b)  \nonumber \\
\propto p(Z_t =k, z_{-t,-(t+1)},z_{t+1}=l | \beta ; \alpha, \bar a, \bar b)  f_k(Y_t)
\end{gather}
Where $f_k(Y_t)=p(Y_t , Y_{-t}| Z_t=k,Z_{-t}; \bar a, \bar b)$. The first term can be evaluated to the following by integrating out $\pi$ as 
\begin{gather}
\label{sup:samplez}
p(Z_t =k | Z_{-t,-(t+1)},Z_{t+1}=l , \beta ; \alpha) = \nonumber  \\
\begin{cases}
   (n^{-t}_{z_{t-1},k} + \alpha \beta_k) \frac{\alpha \beta_l +  (n_{k,l}^{-(t)} + \delta(Z_{t-1},k) \delta(k,l)) }{ \alpha + n_{k,.}^{-(t)} + \delta(Z_{t-1},k)} & \text{k $\in [K]$}  \\
   (\alpha \beta_{K+1}) \frac{ \alpha \beta_l) }{ ( \alpha) }  & \text{k=K+1 }   
  \end{cases}
\end{gather}
Where $n_{k,l}^{-t}=\sum_{\bar t \neq t, \bar t \neq t+1} \delta(Z_{\bar t}, k) \delta(Z_{\bar t+1}, l) $.
The second term $f_k(Y_t)$ is obtained from the equation \ref{sup:zyterm}.

\subsection{Inference : Sparse-HMM-DP-MMVP: }
\label{sup:inference_Sparse-HMM-DP-MMVP}
Sparse-HMM-DP-MMVP Inference is computationally less expensive due to the selective modeling of covariance structure.
However, inference for Sparse-HMM-DP-MVPM requires sampling of $b_{k,j} ,  j \in [M], k=1, \hdots $ in addition to 
latent variables in section \ref{sup:inference-HMM-DP-MMVP} introducing challenges in the non-parametric setting that we 
discuss in this section. Note: Variables, $\eta, \Lambda$ and $\hat \lambda$  are collapsed for faster mixing. 


\underline{\textbf{ Update for $b_{k,j}$ } }:
The update can be written as a product:
\begin{equation}
\label{sup:bupdate}
p(b_{k,j} | b_{-k,j}, Y, Z ) \sim p(b_{k,j} | b_{-k,j}) p(Y| b_{k,j}, b_{-k,j}, Z ; \bar a \bar b)
\end{equation}
By integrating out $\eta$, we simplify the first term as follows 
where $c_j^{-k}=\sum_{\bar k \neq k, \bar k=1}^K b_{k,j}$ is the  number of clusters (excluding k) with dimension $j$ active.
\[p(b_{k,j} | b_{-k,j}) \propto \frac{c_j^{-k} + b_{k,j} + a' -1}{K + a' + b' -1}\] 


The second term can be simplified as follows in terms of $F_{k,j,l}$ as defined in equation \ref{sup:fkjl} by collapsing the $\Lambda$ variables
and $\hat F_{j}$ obtained from integrating out the $\hat \lambda$ variables.
\begin{gather} \label{sup:Sparse_yall_update}
p(Y| b_{k,j}, b_{-k,j}, Z ; \bar a \bar b) \propto \underset {  j \leq l \in [M]} {\Pi} F_{k,j,l} ^{b_{k,j} b_{k,l}} \underset { j \in [M]} {\Pi} \hat F_{j} \\
\label{sup:hatfkjl}
\text{Where } \hat F_{j}= \frac{\Gamma(\hat a+\hat S_{j})}{(\hat b+\hat n_j)^{(\hat a+\hat S_{j})} \underset{t,j: b_{Z_t,j}=0} {\Pi} Y_{t,j,j}!}
\end{gather}. 
And $\hat S_{j}=\sum_t  Y_{t,j,j} (1-b_{Z_t,j})$  and $\hat n_{j}=\sum_t  (1-b_{Z_t,j})$ 

\underline{\textbf{ Update for $Z_t$} }:
Let ${\bf b^o}=\{ {\bf b_k} : k \in [K]\}$ be the variables selecting active dimensions for the existing clusters. 
The update for cluster assignments $Z_t$, $t \in [T]$ while similar to the direct assignment sampling algorithm of HDP\cite{HDP}, 
has to handle the case of evaluating the probability of creating a new cluster with an unknown $b_{k+1}$ .

The conditional for $Z_t$ can be written as a product of two terms as that in equation \ref{sup:updatez}
 \begin{gather}
 \label{sup:inf:sparse:samplez}
 p(Z_t =k | Z_{-t,-(t+1)},z_{t+1}=l, {\bf b^{old}}, X, \beta , Y; \alpha, \bar a, \bar b)  \nonumber  \\
\propto p(Z_t =k, Z_{-t,-(t+1)},_{t+1}=l | \beta ; \alpha, \bar a, \bar b) \nonumber \\
 p(Y_t | Z_t=k,Z_{-t}, Y_{-t}, \bf b^{old}; \bar a, \bar b)   
 \end{gather}
The first term can be simplified in a way similar to \cite{StickyHDPHMM}. To evaluate the second term, two cases need to be considered. 

{\bf Existing topic ($k  \in [K]$) :} 
In this case, the second term $p(Y_t | {\bf b^{o}} Z_t=k,Z_{-t}, Y_{-t}; \bar a, \bar b)$  can be simplified by 
integrating out the $\Lambda$ variables as in equation \eqref{sup:Sparse_yall_update}.

{\bf New topic ($k = K+1$) :} 
In this case, we wish to compute $p(Y_t | {\bf b^{o}}, Z_t=K+1,Z_{-t}, Y_{-t}; \bar a, \bar b)$. Since this expression is not conditioned on $b_{K+1}$, evaluation of this 
term requires summing out $b_{K+1}$ as follows.
\begin{align*}
p(Y_t | {\bf b^{o}}, Z_t=K+1,Z_{-t}, Y_{-t}; \bar a, \bar b) = & \\
\sum_{b_{K+1}} p(b_{K+1} | {\bf b^o}, \eta) p(Y_t |{\bf  b^{o}}, b_{K+1},  Z_t=K+1,Z_{-t}, Y_{-t}; \bar a, \bar b)
\end{align*}

Evaluating this summation involves exponential complexity. Hence we resort to a simple numerical approximation as follows. 
Let us denote $p(Y_t | {\bf b^{o}}, b_{K+1},  Z_t=K+1,Z_{-t}, Y_{-t}; \bar a, \bar b)$ as $h(b_{K+1})$
The above expression can be viewed as an expectation  $E_{b_{K+1}}[ h(b_{K+1}) | {\bf b^o}]$. 
and can be approximated numerically by drawing samples of $b_{K+1}$ with probability $p(b_{K+1} | {\bf b^o})$. We use Metropolis Hastings algorithm
to get a fixed number S of samples using the proposal distribution that flips each element of b independently with a small probability $\hat p$.
The intuition here is that we expect the feature selection vector for new cluster, $b_{K+1}$ to be reasonably close to $b_{Z^{old}_t}$, 
the selection vector corresponding to the previous cluster assignment for this data point. 
In our experiments we set S=20 and $\hat p$=0.2 to give reasonable results.
 
We note that in \cite{SparseTM}, the authors address a similar problem of feature 
selection, however in a multinomial DP-mixture setting, by collapsing the $b$ selection variable. However, their technique is specific to 
sparse Multinomial DP-mixtures. 

\underline{\textbf{ Update for $Y_{t,j,l}$}}:
The update for $Y_{t,j,l}$ is similar to that in section \ref{sup:inference_Y_DP-MMVP} with the following difference.
We sample only $\{Y_{t,j,l} : b_j=1, b_l=1\}$ and 
the rest of the elements of Y are set to 0 with the exception of diagonal elements for the inactive dimensions.
We note that for the inactive dimensions $\{j : j  \in [M] , b_{Z_t,j}=0\}$, the value of $X_{t,j}=Y_{t,j,j}$ and hence 
can be set directly from the observed data without sampling.

For the active dimensions, $\{Y_{t,j,l} : b_j=1, b_l=1, j \leq l \in [M]\}$ we sample using a procedure 
similar to that in section \ref{sup:inference_Y_DP-MMVP} by sampling $Y_{t,j,l}, j \neq l$ to preserve 
the constraint $X_{t,j}=\sum_{l=1}^M Y_{t,j,l}$, restricting the support of the random variable in a procedure 
similar to section \ref{sup:inference_Y_DP-MMVP}. 
\begin{align}
\label{sup:inf:sparse:yupdate}
p(Y_{t,j,l} | Y_{-t,j,l}, Z, \bar a, \bar b, \hat a, \hat b) \propto 
 F_{k,j,l} F_{k,j,j} F_{k,l,l}  \underset { 1<=j<=M} {\Pi} \hat F_{j} 
\end{align}

\begin{algorithm}[htp]
\Repeat{convergence}{
\For {$t=1, \hdots , T$} {
  Sample $Z_t$ from Eqn \ref{sup:inf:sparse:samplez} (Alt: Eqn \ref{sup:updatez})  \\
  \For{$ j \leq l \in [M] $}{
     \If{$b_{Z_t,j}=b_{Z_t,k}=1$}{
         Sample $Y_{t,j,l}$ from Eqn \ref{sup:inf:sparse:yupdate} (Alt: Eqn \ref{sup:yupdate})\\
         Set $Y_{t,j,j}=X_{t,j}-\sum_{\bar j=1}^M Y_{t,j,\bar j}$ \\
         Set $Y_{t,l,l}=X_{t,l}-\sum_{\bar l=1}^M Y_{t,l,\bar l}$ 
     }
  }  
}
\For{$j=1, \hdots, M, k=1, \hdots, K$}{
  Sample $b_{k,j}$ from Eqn \ref{sup:bupdate} (Alt: Set $b_{k,j}=1_M$)
}
 \For{$k=K, \hdots, M, k=1, \hdots$}{
   $m_k=0$ \\
   \For {$i=1, \hdots , n_k$} {
   $u \sim Ber(\frac{\alpha \beta_k}{i+\alpha \beta_k})$, if $(u==1)m_k++$ \\
   }
 }
 $[\beta_1 \beta_2 \hdots \beta_K \beta_{K+1}]  | m, \gamma \sim Dir(m_1, \hdots , m_k, \gamma)$
 }
 \caption{Inference: Sparse-HMM-DP-MMVP Inference steps(The steps for HMM-DP-MMVP are similar and are shown as 
 alternate updates in brackets)}
 \label{sup:alg:inference}
\end{algorithm}



%


\vspace{3em}
\section*{\fontsize{14}{15}\selectfont Appendix C: Experiment Details}
 \addtocounter{section}{1}
\subsection{Dataset Details: }
We perform experiments on publicly available real world block I/O traces from enterprise servers at Microsoft 
Research Cambridge \cite{MSFTdataset}. They represent diverse enterprise workloads.
These are about $36$ traces comprising about a week worth of data,  thus allowing us to study long ranging temporal dependencies. 
We eliminated $26$ traces that are write heavy (write percentage $>25\%$) as we are focused on read cache. See Table ~\ref{sup:tab:data} for the datasets and their read percentages.  We present our results on the remaining $10$ traces. 
We also validated our results on one of our internal workloads, NT1 comprising data collected over 24 hours.    

We divide the available trace into two parts $\mathcal D^{lr}$ that is aggregated into $T_{lr}$ 
count vectors and $\mathcal D^{op}$ that is aggregated into $T_{op}$ count vectors.
We use a split of 50\% data for learning phase and 50\% for operation phase for our experiments such that $T_{lr}=T_{op}$.

 \begin{table}[htbp]
 \centering
 \caption{Dataset Description}
 \resizebox{8cm}{!} {
 \begin{tabular}{| l | c|r|r|}\hline
  Acro    & Trace &Description&Rd   \\ 
  -nym  & Name &     &\% \\ \hline
  MT1&CAMRESWEBA03-lvm2   &   Web/SQL Srv  &99.3  \\
  MT2&CAMRESSDPA03-lvm1   &   Source control  &97.9  \\
  MT3&CAMRESWMSA03-lvm1   &   Media Srv    &92.9  \\
  MT4&CAM-02-SRV-lvm1     &    User files     &89.4  \\
  MT5&CAM-USP-01-lvm1     &    Print Srv   &75.3  \\
  MT6&CAM-01-SRV-lvm2     &    User files     &81.1  \\
  MT7&CAM-02-SRV-lvm4     &    User files     &98.5  \\
  MT8&CAMRESSHMA-01-lvm1&      HW Monitoring &95.3  \\                                        
  MT9&CAM-02-SRV-lvm3    &     project files  &94.8  \\
  MT10&CAM-02-SRV-lvm2     &   User files     &87.6  \\ \hline
 NT1 & InHouse Trace &          Industrial &95.0 \\  \hline
\end{tabular}
}
\label{sup:tab:data}
\end{table}
\subsection{Design of Simulator: }
The design of our baseline simulator and that with preloading is described below.

{\bf Baseline: LRU Cache Simulator}: We build a cache simulator that services access requests from the trace maintaining a cache. 
When a request for a new block comes in, the simulator checks the cache first. If the block is already in the cache it records a hit, else it records a miss and adds this block 
to the cache. The cache has a limited size (fixed to 5\% the total trace size). When the cache is full, and a new block is to be added
to the cache, the LRU replacement policy is used to select an existing block to remove. 
We use the hitrates obtained by running the traces on this simulator as our baseline. 

{\bf LRU Cache Simulator with Preloading}: In this augmented simulator, at the end of every $\nu =30s$, predictions are made using the 
framework described in Section \ref{sec:frame} and loaded into the cache (evicting existing blocks based on the LRU policy as necessary). 
While running the trace, hits and misses are kept track of, similar to the previous setup. The cache size used is the same as 
that in the previous setting.

\subsection{Hitrate Values: }
Figure \ref{hitrates} in our paper shows a barchart of hitrates for comparison.
In table \ref{hitrate_table} of this section, the actual hitrate values are provided comparing preloading with 
Sparse-HMM-DP-MMVP and that with baseline LRU simulator without preloading. We note that we show a dramatic improvement in 
hitrate in 4 of the traces while we beat the baseline without preloading in most of the other traces.

\begin{table}[htp]\footnotesize
\centering
\begin{tabular}{| l |  r| r|r|r|}\hline
 Trace     & Preloading&LRU \\
  &   Sparse-HMM-  &Without   \\
  &    DP-MMVP &Preloading \\\hline 
  MT1 &  {\bf 52.45 \% } &  0.02  \% \\
  MT2 &  {\bf 23.97 \% } &  0.1   \% \\
  NT1 &  {\bf 34.03 \% } &  03.66 \% \\ 
  MT3 &  {\bf 34.40 \% } & 6.90   \% \\
  MT4 &  {\bf 52.96 \% } &  42.12 \% \\
  MT5 &  {\bf 40.15 \% } &  39.75 \% \\   
  MT6 &  {\bf 3.80  \% } &  3.60  \% \\ 
  MT7 &  {\bf 5.44  \% } &  5.22  \% \\ 
  MT8 &  {\bf 98.15 \% } & 98.04  \% \\ 
  MT9 &  {\bf 65.54 \% } &  65.54 \%  \\
  MT10&  {\bf 0.0   \% } &  0.0   \% \\\hline
\end{tabular}      
\begin{centering}\caption{Comparing the hitrate with preloading with HMM-DP-MMVP and Sparse-HMM-DP-MMVP with hitrate of 
LRU without preloading. We note that Sparse-HMM-DP-MMVP beat the baseline showing dramatic improvement for 
the first four traces, and do well in comparison with the baseline for all the remaining traces.}
\label{hitrate_table}
\end{centering}
\end{table}


\subsection{Effect of Training Data Size: } 
We expect to capture long range dependencies in access patterns when we observe a sufficient portion of the trace for training 
where such dependencies manifest. We show this by running our algorithm for different splits of train and test data 
(corresponding to the learning phase and the operational phase) for NT1 trace.

We observe that when we see at least 50\% of the trace, there is a marked improvement in hitrate for the NT1 trace. 
Hence we use 50\% as our data for training for our experimentation.

In a real world setting, we expect the amount of data required for training to vary across workloads. To adapt our 
methodology in such a setting periodic retraining to update the model with more and more data for learning as it is 
available is required. Exploring an online version of our models might also prove useful in such settings.

\begin{figure}[htbp]
\label{sup:trainingsize}
\centering \includegraphics[height=1.20in,width=3.5in]{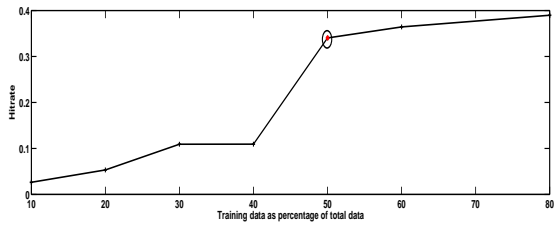}
\centering \caption{\small \sl Hitrate with increasing percentage of training data}  
\label{sup:spatiotemporal}
 \vspace{-1em}
\end{figure}  


\end{document}